\documentclass[aps,preprint]{revtex4}%
\usepackage{amsfonts}
\usepackage{amsmath}
\usepackage{amssymb}
\usepackage{graphicx}
\usepackage{pdfpages}
\usepackage{xcolor}
\usepackage{caption}
\usepackage{lineno,hyperref}
\usepackage{changes}%
\hypersetup{colorlinks =true, allcolors = blue}
\providecommand{\U}[1]{\protect\rule{.1in}{.1in}}

\begin{document}
\preprint{HEP/123-qed}
\title{GUP modified Hawking Radiation and Transmission/Reflection Coefficients of Rotating Polytropic
Black Hole}
\author{Sara Kanzi}
\affiliation{}
\author{}
\affiliation{}
\author{\.{I}zzet Sakall{\i}}
\affiliation{Physics Department, Eastern Mediterranean
University, Famagusta 99628, North Cyprus via Mersin 10, Turkey.}
\author{}
\affiliation{}

\begin{abstract}
In this paper, we study the GUP (Generalized Uncertainty Principle) modified Hawking radiation of rotating polytropic black hole given in the Boyer–Lindquist coordinates. To this end, the GUP modified Klein-Gordon equation is applied for investigating  the quantum tunneling of scalar particles from the polytropic black hole. After reducing the obtained radial wave equation to the one-dimensional Schr\"{o}dinger equation, we derive the reflection and transmission probabilities of the radiation. A detail discussion on how the reflection and transmission (greybody factor) probabilities are derived for this black hole is given. The results are graphically depicted and the relevant physical interpretations are made.

\end{abstract}
\volumeyear{ }
\eid{ }
\date{\today}
\received{}

\maketitle
\tableofcontents

\section{Introduction}%

Black holes are usually surrounded by different matters, such as accretion
disks, galactic nuclei, strong magnetic fields, other stars and planets etc. Therefore, interactions happen with their
surroundings and affect to the properties and stability of the black hole
\cite{N1}. Moreover, a black hole even interacts with the vacuum surrounding it. A pair of virtual particles appear right on the edge of the event horizon of a black hole. Therefore, one particle falls in to the black hole while its counterpart escapes, thus appearing as if the black hole emitted a particle and lost mass in the process. Namely, a black hole can 'evaporate' over huge time spans via a process the so-called "Hawking radiation". So, a real black hole can never be fully described by its basic parameters
and therefore it is always in the perturbed state. 

Coalescing binary neutron stars is one of the most promising sources of gravitational waves. The coalescence of two identical neutron stars are represented by polytropes, which were considered first time by
Nakamura and Oohara in 1989 \cite{N2}. They performed the post-Newtonian (in the context of General Relativity) three dimensional simulations including the radiation reaction of gravitational waves to know the amplitude and the wave form of gravitational waves at the coalescing events, which are considered as astrophysical problems
\cite{N3,N4,N5,N6}. Recently, black hole solutions to the
polytropic equations of state (PEoS) have been obtained in Ref. \cite{N7}.  PEoS, which are considered in various astrophysical situations \cite{N8,N9}, are defined as follows
\begin{equation}
P=K\rho^{1+\frac{1}{n}}, \label{N1}%
\end{equation}
where $n$ is polytropic index and $K$ denotes the positive constant. Moreover, it is worth noting that PEoS can be considered in a cosmological context to model
the matter content \cite{Nd6}. In another important study, a rotating black hole solution is derived from a
spherically symmetric and static polytropic black hole by applying the Newman-Janis algorithm without complexification \cite{S1}.

Based on the classical theories, black holes do not radiate and their
temperatures are considered to be zero. In 1974, Hawking proved that black holes can radiate with a non-zero characteristic temperature \cite{K1}. Hawking radiation is nothing but the quantum tunneling of virtual particles buried in the vacuum around the event horizon of the black hole \cite{2}. Hawking radiation can be regarded as the onset of quantum gravity theory \cite{3,4,5}. GUP is consistent with various proposals of quantum gravity such as string theory, loop quantum gravity, doubly special relativity, and predicts both a minimal length uncertainty and a maximal observable momentum \cite{6,7,8,9,10}. Since the magnificent discovery of Hawking radiation \cite{1}, many valuable studies have taken their place in the current literature (see \cite{S10,S11,S12,S13,S14} and references therein) and continue to do so. 

GUP is the modified version of the standard uncertainty principle:
\begin{equation}
\Delta x\Delta p\geq\frac{\hbar}{2}\left(  1+\beta\Delta p^{2}\right)  ,
\label{N2}%
\end{equation}

where $\beta=\frac{\beta_{0}l_{p}^{2}}{\hbar^{2}},$ $l_{p}$ is the Planck length, and $\beta_{0}$ is a constant. Today, GUP is subject to many interesting physical problems and as a result, many academic articles are published for this cause \cite{11,12,13,14}. Few specific examples of these studies are as follows: GUP modified Hawking radiations of $2+1$ dimensional Martinez-Zanelli black hole \cite{12}, the warped DGP gravity black hole \cite{15}, Schwarzschild-like black hole in bumblebee gravity model \cite{16} and  Kerr-type black holes \cite{17,18,19,20}.

Curvature in the spacetime naturally alters the measured Hawking radiation depending on the location. This modification in the Hawking radiation is determined by the greybody factor \cite{S20,S21}. Namely, the radiation spectrum at the horizon is different from the spectrum recorded by an observer at spatial infinity. The emission rate, which is determined by a faraway observer is written by \cite{21}
\begin{equation}
\gamma\left(  \omega\right)  =\left(  \frac{\left\vert A_{l,m}\right\vert
^{2}d^{3}k}{\left(  e^{\frac{\omega}{T_{H}}}\pm1\right)  \left(  2\pi\right)
^{3}}\right)  , \label{N3}%
\end{equation}
where $\left\vert A_{l,m}\right\vert ^{2}$ is the greybody factor, $T_{H}$ represents the Hawking temperature, and $\left(  \pm\right)  $ signs stand for the fermions and bosons, respectively. In fact, the greybody factor (rate of absorption probability which is a frequency-dependent quantity) is nothing else than the transmission amplitude as the field modes propagate from the horizon region to spatial infinity by passing the relevant effective potential, created due to the spacetime geometry \cite{22,23}. Therefore, the one part of scattered Hawking radiation
is reflected back into the black hole, while the other part is transmitted to spatial infinity. The transmission probability in this context is also known as the greybody factor. There are different methods to compute the greybody factor. Among them the well-known ones are the WKB approximation method \cite{24,S24,D24}, the rigorous bound method \cite{25,26,29}, and the matching technique \cite{27,28}. Some of the other
methods can be seen in \cite{30,31} and the references therein.

This paper is organized as follows: In Sec. \ref{sec2}, we briefly introduce the rotating polytropic black hole. In Sect. \ref{sec3}, we study the Hawking radiation of spin-0 particles, within the framework of the GUP, emitted from the rotating polytropic black hole. To this end, in Sect. \ref{sec4}, we employ the GUP modified Klein-Gordon equation together with the dragging coordinates. In Sec. \ref{sec5}, we compute the reflection and transmission probabilities by
using the WKB method. Section \ref{sec6} is devoted to summary and conclusions. We follow the metric convention $(-,+,+,+)$.

\section{Rotating Polytropic Black Hole}\label{sec2}

In this section, we shall briefly represent the physical properties of the polytropic black hole whose metric is described in the Kerr coordinates as follows%

\begin{equation}
ds^{2}=\left(  1-\frac{2f}{\rho^{2}}\right)  dt^{2}-\frac{\rho^{2}}{\Delta
}dr^{2}+\frac{4af\sin^{2}\theta}{\rho^{2}}dtd\phi-\rho^{2}d\theta^{2}%
-\frac{\Sigma\sin^{2}\theta}{\rho^{2}}d\phi^{2}, \label{1}%
\end{equation}
where 
\begin{equation}
 \begin{aligned}
2f=r^{2}\left(  1-F\right),\\
\rho^{2}=r^{2}+a^{2}\cos^{2}\theta, \quad \\
\Sigma=\left(  r^{2}+a^{2}\right)  ^{2}-a^{2}\Delta\sin^{2}\theta,
\end{aligned}
\end{equation}
in which%
\begin{equation}
\Delta=a^{2}+r^{2}F, \label{2}%
\end{equation}
and%
\begin{equation}
F=\frac{r^{2}}{L^{2}}-\frac{2M}{r}. \label{3}%
\end{equation}
In Eq. (\ref{3}), $L^{2}=-\frac{3}{\Lambda}$ where $\Lambda$ represents the cosmological
constant. Throughout our study, we shall consider the negative cosmological constant, where empty space itself has negative energy density but positive pressure, like the anti-de Sitter space. The solution of $\Delta\left(  r_{\pm}\right)  =0$ yields
the horizons of the black hole. Namely, we have
\begin{equation}
a^{2}+r_{\pm}^{2}F(r_{\pm})=0,    
\end{equation}
which yields 
\begin{multline}
r_{+}=\frac{\left[  -\frac{2\sqrt[3]{6}a^{2}L^{2}}{d\left(  a,M,L\right)
}+\frac{6L^{2}M}{\sqrt{\frac{\sqrt[3]{6}a^{2}L^{2}}{d\left(  a,M,L\right)
}+\frac{1}{2}d\left(  a,M,L\right)  }}-d\left(  a,M,L\right)  \right]  ^{1/2}%
}{2^{5/6}\sqrt[3]{3}}\\
+\frac{\left[  \frac{2\sqrt[3]{6}a^{2}L^{2}}{d\left(  a,M,L\right)  }+d\left(
a,M,L\right)  \right]  ^{1/2}}{2^{5/6}\sqrt[3]{3}}, \label{S5}%
\end{multline}
where
\begin{equation}
d\left(  a,M,L\right)  =\sqrt[3]{\sqrt{3}\sqrt{27L^{8}M^{4}-16a^{6}L^{6}%
}+9L^{4}M^{2}}. \label{6}%
\end{equation}

One can easily observe from the above horizon condition that there exists a bound on the spin/mass-parameter, $\frac{a}{M}$. The allowed values for the spin/mass-parameter are constrained as follows \cite{S1}:
\begin{equation}
\frac{a}{M}<3^{1 /2}\left(\frac{L}{4M}\right)^{1 / 3}. \label{is9}
\end{equation}

It is worth noting that above result \ref{is9} differs from the Kerr case in which the constraint corresponds to $\frac{a}{M}<1$. The Hawking temperature for the rotating polytropic black hole is given by \cite{S1,S2},%
\begin{equation}
T_{H}=\frac{1}{4 \pi} \lim _{r \rightarrow r_{+}} \frac{\partial_{r} g_{t t}}{\sqrt{g_{t t} g_{r r}}}=\frac{-a^{2}\left(  L^{2}\left(  M+r_{+}\right)  -2r_{+}^{3}\right)
+L^{2}Mr_{+}^{2}+r_{+}^{5}}{2\pi L^{2}\left(  a^{2}+r_{+}^{2}\right)  ^{2}}.
\label{4}%
\end{equation}

\section{GUP Modified Hawking Radiation of Rotating Polytropic Black Hole}\label{sec3}

In this section we first derive the diagonal form (without cross-term) of the stationary metric \eqref{1} by applying the frame-dragging effect. In the framework of general relativity, the frame-drag effect was first introduced in 1918 by the Austrian physicists J. Lense and H. Thirring \cite{SS2}. That is why this application is also known as the Lense–Thirring effect, which states that spacetime will churn around a massive, rotating body. Put simply, the rotating mass “drags along” spacetime in the vicinity. 

An observer rests in the dragging coordinate system, which behaves like a kind of locally non-rotating coordinate system, would suppose himself in a static coordinate system since he co-rotates with the rotating black hole. Namely, the observer rotates with the angular velocity of
$\Omega=-\frac{g_{03}}{g33}$ \cite{S3,S4}. Performing the dragging coordinate transformation \cite{S4} to the metric \eqref{1}, one can get%
\begin{equation}
ds^{2}=-\frac{\Delta\rho^{2}}{\Sigma}dt^{2}+\frac{\rho^{2}}{\Delta}dr^{2}%
+\rho^{2}d\theta^{2}=-Fdt^{2}+\frac{1}{G}dr^{2}+Kd\theta^{2}, \label{7}%
\end{equation}

so that 
\begin{equation}
 \begin{aligned}
F=\frac{\Delta\rho^{2}}{\Sigma},\\
G=\frac{\Delta}{\rho^2},\\
K=\rho^2.
\label{7n}%
\end{aligned}
\end{equation}

The generalized Klein-Gordon equation with GUP is given by \cite{SS5}%
\begin{equation}
-\left(  i\hbar\right)  ^{2}\partial^{t}\partial_{t}\psi=\left[  \left(
-i\hbar\right)  ^{2}\partial^{i}\partial_{i}+m^{2}\right]  \left\{
1-2\beta\left[  \left(  -i\hbar\right)  ^{2}\partial^{i}\partial_{i}%
+m^{2}\right]  \right\}  . \label{8}%
\end{equation}
By using a standard ansatz for the scalar wave $\psi$:
\begin{equation}
\psi=\exp\left[  \frac{i}{\hbar}I\left(  t,r,\theta\right)  \right]  ,
\label{9}%
\end{equation}
one can get
\begin{equation}
\left(  \frac{1}{F}\right)  \left(  \partial_{t}I\right)  ^{2}=\left[
G\left(  \partial_{r}I\right)  ^{2}+\frac{1}{K}\left(  \partial_{\theta
}I\right)  ^{2}+m^{2}\right]  -2\beta\left[  G\left(  \partial_{r}I\right)
^{2}+\frac{1}{K}\left(  \partial_{\theta}I\right)  ^{2}+m^{2}\right]  ^{2}.
\label{10}%
\end{equation}

To solve the above equation, we carry out the separation of variables technique as follows%
\begin{equation}
I=-\left(  \omega-j\Omega\right)  t+R\left(  r\right)  +W\left(
\theta\right)+\mathcal{K}, \label{11}%
\end{equation}
where $\omega$ and $j$ are the energy of the particles and the angular momentum, respectively. $\mathcal{K}$ is a complex constant. After substituting Eq. \eqref{11} into Eq. \eqref{10}, we get%
\begin{equation}
a\left(  \partial_{r}R\right)  ^{4}+b\left(  \partial_{r}R\right)  ^{2}+c=0,
\label{12}%
\end{equation}
where%
\begin{align}
a  &  =-2\beta G^{2},\nonumber\\
b  &  =G\left(  1-4\beta m^{2}\right) \nonumber\\
c  &  =\frac{1}{K}\left(  \partial_{\theta}W\right)  ^{2}+m^{2}-2\beta
m^{4}-\frac{1}{F}\left(  \omega-j\Omega\right)  ^{2}. \label{S13}%
\end{align}

The integral solution of Eq. \eqref{12} at the horizon is found as follows%
\begin{multline}
R_{\pm}=\left.  \pm\int\frac{dr}{\sqrt{FG}}\sqrt{\left(  \omega-j\Omega
\right)  ^{2}-F\left(  \frac{1}{\rho^{2}\left(  r_{h}\right)  }\left(
\partial_{\theta}W\right)  ^{2}+m^{2}\right)  }\right. \\
\left.  \times\left[  1+\beta\left(  m^{2}+\frac{1}{\rho^{2}\left(
r_{h}\right)  }\left(  \partial_{\theta}W\right)  ^{2}\right)  +\beta
\frac{\left(  \omega-j\Omega\right)  ^{2}}{F}\right]  ,\right.  \label{S14}%
\end{multline}
where $+/-$ represent the outgoing and ingoing solutions, respectively. On the other hand, near the event horizon (r$\rightarrow r_{+}$), $\Delta\left(  r\right)$ function behaves as
\begin{equation}
\Delta\left(  r\right)  =\Delta\left(  r_{h}\right)  +\left(  r-r_{h}\right)
\Delta_{,r}\left(  r_{h}\right)  +O\left(  \left(  r-r_{+}\right)
^{2}\right)  \simeq\left(  r-r_{h}\right)  \Delta_{,r}\left(  r_{h}\right),
\label{15}%
\end{equation}

which enables us to integrate Eq. \eqref{S14} near the event horizon. Using the residue theorem, we obtain%
\begin{equation}
R_{\pm}=\pm\frac{i\pi\left(  \omega-j\Omega_{h}\right)  \left(  r_{h}^{2}%
+a^{2}\right)  ^{2}L^{2}}{2\left(  r_{h}^{5}+Mr_{h}^{2}L^{2}+2r^{3}%
a^{2}-ML^{2}a^{2}-r_{h}a^{2}L^{2}\right)  }\left[  1+\beta\Xi\right]  ,
\label{16}%
\end{equation}
in which%
\begin{multline}
\Xi=\frac{1}{2}\left(  m^{2}+\frac{\left(  \frac{dw}{d\theta}\right)  ^{2}%
}{r_{h}^{2}+a^{2}\cos^{2}\theta}-\right. \\
\left.  \frac{\left(  \omega-j\Omega_{h}\right)  ^{2}\left(  r_{h}^{2}%
+a^{2}\right)  ^{4}}{3\left(  r_{h}^{2}+a^{2}\cos^{2}\theta\right)  \left(
r_{h}^{5}+Mr_{h}^{2}L^{2}+2r_{h}^{3}a^{2}-ML^{2}a^{2}-r_{h}a^{2}L^{2}\right)
^{2}}\right)  . \label{17}%
\end{multline}

According to the WKB approximation, the emission and absorption tunneling probabilities
of the scalar particles performing the quantum tunneling through the event horizon are defined as follows \cite{SS6},%
\begin{equation}
\Gamma\left(  emission\right)\propto \exp \left(-2 \operatorname{Im}I\right)  =\exp\left[  -2\left(  \operatorname{Im}%
R_{+}+\operatorname{Im}\mathcal{K}\right)  \right]  , \label{18}%
\end{equation}
and%
\begin{equation}
\Gamma\left(  absorption\right) \propto \exp \left(-2 \operatorname{Im}I\right) =\exp\left[  -2\left(  \operatorname{Im}%
R_{-}+\operatorname{Im}\mathcal{K}\right)  \right].  \label{19}%
\end{equation}
According to the classical approach, the probability of any incoming particles crossing the event horizon has a $100 \%$ chance for entering the black hole. Therefore, it is necessary to set
\begin{equation}
\operatorname{Im\mathcal{K}}=-\operatorname{Im}R_{-}    
\end{equation}
in the above equation. From Eq. \eqref{18}, it is obvious that
\begin{equation}
R_{+}=-R_{-}.
\end{equation}
Thus, the tunneling rate for the scalar particles reads%
\begin{equation}
\Gamma=\exp\left[  -4\operatorname{Im}R_{+}\right]  , \label{20}%
\end{equation}
which gives the following result for the rotating polytropic black hole:%
\begin{equation}
\Gamma=\exp\left[  \frac{-2\pi\left(  \omega-j\Omega_{h}\right)  \left(
r_{h}^{2}+a^{2}\right)  ^{2}L^{2}}{\left(  r_{h}^{5}+Mr_{h}^{2}L^{2}%
+2r^{3}a^{2}-ML^{2}a^{2}-r_{h}a^{2}L^{2}\right)  }\left(  1+\beta\Xi\right)
\right]  . \label{21}%
\end{equation}

This is the probability of an outgoing scalar particle tunneling  from the event horizon of the rotating polytropic black hole. If one compares Eq. \eqref{21} with the Boltzmann factor \cite{SS7}, it is seen that the term in the
first bracket of the exponential function is equal to the inverse temperature of the radiating black hole. Ultimately, we obtain the GUP modified Hawking temperature $T_{GUP}$ of the rotating polytropic black hole as follows%
\begin{equation}
T_{GUP}=\frac{T_{H}}{1+\beta\Xi}. \label{22}%
\end{equation}
Recall that $T_{H}$ is the standard Hawking temperature of the rotating polytropic black hole; see Eq. \eqref{4}.

\section{Effective Potential for bosons propagating in rotating Polytropic Black Hole Geometry}\label{sec4}

In this section, we shall derive the effective potential felt by the scalar particles while moving in the rotating polytropic black hole spacetime. To this end, we consider a test particle with spin-0 having mass $\mu_{0}$ and let it moves in the geometry of the rotating polytropic black hole metric \eqref{1}. The wave
equation of the corresponding scalar field is derived by the general Klein-Gordon equation:%
\begin{equation}
\left(  \nabla^{\nu}\nabla_{\nu}-\mu_{0}^{2}\right)  \Psi=0, \label{s22}%
\end{equation}
where $\nabla_{\nu}$ is the covariant derivative. Therefore, Eq. \eqref{s22} can be also
written as
\begin{equation}
\left[  \frac{1}{\sqrt{-g}}\partial_{\mu}\left(  g^{\mu\nu}\sqrt{-g}%
\partial_{\nu}\right)  \right]  \Psi=\mu_{0}^{2}\Psi. \label{23}%
\end{equation}

Since the considered metric is Eq. \eqref{1}, one can compute the determinant of its metric tensor as follows
\begin{equation}
g\equiv\det g_{\mu\nu}=-\rho^{4}\sin^{2}\theta. \label{24}%
\end{equation}
After substituting Eq. \eqref{24} into Eq. \eqref{23}, we get
\begin{multline}
\left.  \frac{\Sigma}{\rho^{2}\Delta}\partial_{t}^{2}\Psi+\frac{2af}{\rho
^{2}\Delta}\partial_{t}\partial_{\varphi}\Psi-\frac{1}{\rho^{2}\sin\theta
}\left\{  \cos\theta\partial_{\theta}+\sin\theta\partial_{\theta}^{2}\right\}
\Psi-\frac{1}{\rho^{2}}\partial_{r}\left(  \Delta\partial_{r}\right)
\Psi+\right. \\
\left.  \frac{2af}{\rho^{2}\Delta}\partial_{\varphi}\partial_{t}\Psi
-\frac{\Delta-a^{2}\sin^{2}\theta}{\Delta\rho^{2}\sin^{2}\theta}%
\partial_{\varphi}^{2}\Psi=\mu_{0}^{2}\Psi.\right.  \label{25}%
\end{multline}

We consider an ansatz of the scalar field
\begin{equation}
\Psi\left(  r,t\right)  =R\left(  r\right)  S\left(  \theta\right)
e^{im\varphi}e^{-i\omega t}, \label{26}%
\end{equation}

where $\omega$ represents the frequency of the wave and $m$ is azimuthal quantum number. By using Eqs. \eqref{25} and \eqref{26}, we obtain%
\begin{multline}
\Delta\frac{d}{dr}\left(  \Delta\frac{dR\left(  r\right)  }{dr}\right)  +\\
\left[  m^{2}a^{2}-4afm\omega+\omega^{2}\left(  r^{2}+a^{2}\right)
^{2}-\left(  \omega^{2}a^{2}+\mu_{0}^{2}r^{2}+\lambda_{lm}\right)  \Delta\right]
R\left(  r\right)  =0, \label{s27}%
\end{multline}

where $\lambda_{lm}$ is the eigenvalue of the angular solution whose differential equation is given by%

\begin{equation}
\frac{1}{S\left(  \theta\right)  }\frac{1}{\sin\theta}\frac{d}{d\theta}\left(
\sin\theta\frac{dS\left(  \theta\right)  }{d\theta}\right)  -\frac{m^{2}}%
{\sin^{2}\theta}+c^{2}\cos^{2}\theta=-\lambda_{lm}, \label{28}%
\end{equation}

where $c^{2}=\left(  \omega^{2}-\mu_{0}^{2}\right)  a^{2}.$ Setting the tortoise coordinate \cite{Sd4} as%

\begin{equation}
\frac{dr_{\ast}}{dr}=\frac{r^{2}+a^{2}}{\Delta}. \label{29}%
\end{equation}

\begin{figure}[h]
\centering\includegraphics[width=9cm,height=10cm]{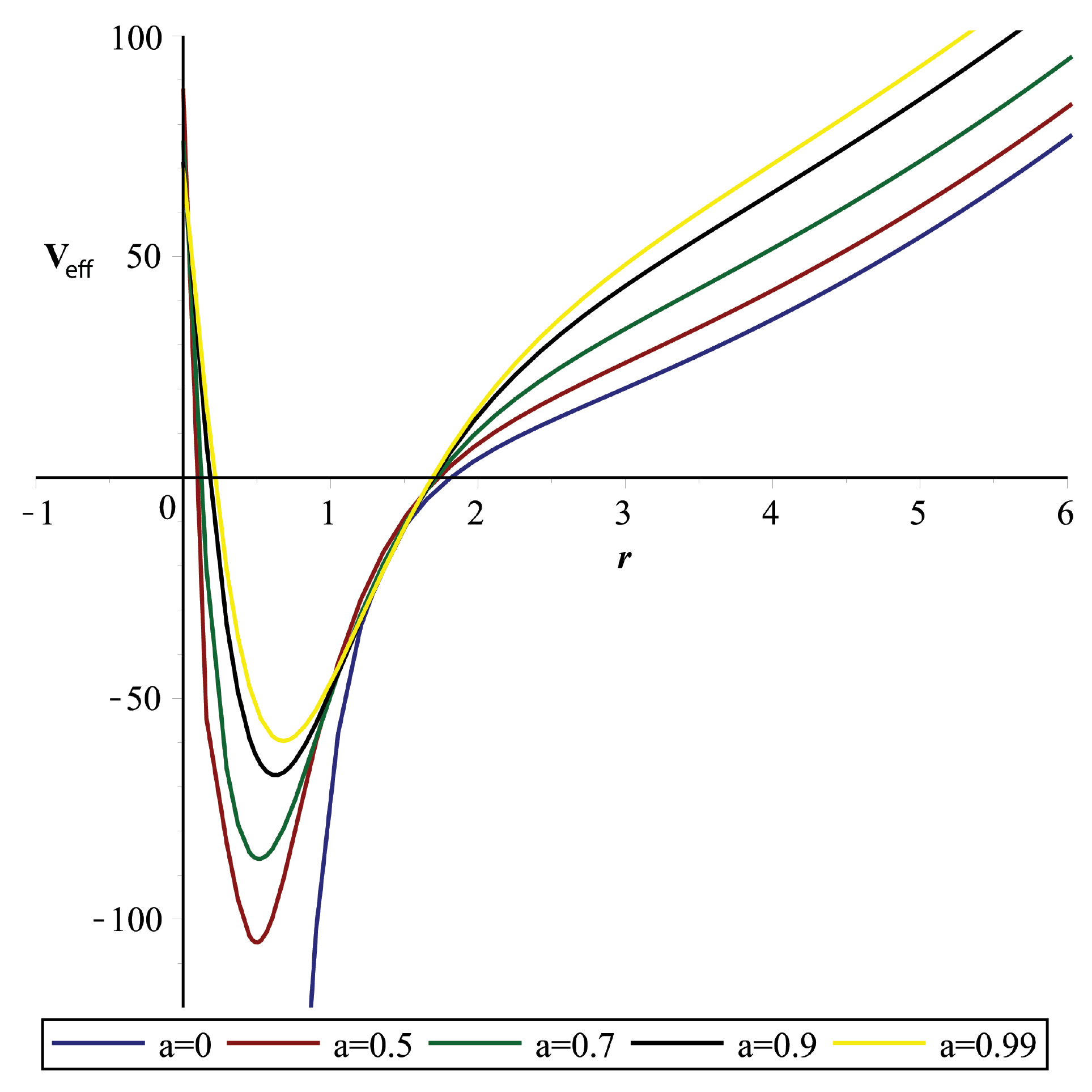}\caption{Plots of
$V_{eff}$ versus $r$ for the spin-0 particles. The physical parameters are
chosen as; $m=1,M=\omega=3, L=1,$ and $\lambda=6$.} \label{fig1}
\end{figure}

The tortoise coordinate $r_{\ast}$ approaches $-\infty$ as $r$ approaches the event radius.  Applying the following transformation%
\begin{equation}
R=\frac{U}{\sqrt{r^{2}+a^{2}}}, \label{30}%
\end{equation}

we get the one dimensional Schr\"{o}dinger-like wave equation:
\begin{equation}
\frac{d^{2}U}{dr_{\ast}^{2}}+\left(  \omega^{2}-V_{eff}\right)  U=0,
\label{31}%
\end{equation}

where the effective potential $V_{eff}$ is found out to be%
\begin{equation}
V_{eff}=\frac{\Delta}{\left(  r^{2}+a^{2}\right)  ^{2}}\left[  \frac
{\Delta^{\prime}r+\Delta}{r^{2}+a^{2}}-\frac{3r^{2}\Delta}{\left(  r^{2}%
+a^{2}\right)  ^{2}}+\frac{4afm\omega-m^{2}a^{2}}{\Delta}+\omega^{2}a^{2}%
+\mu_{0}^{2}r^{2}+\lambda_{lm}\right]  , \label{32}%
\end{equation}
where prime indicates derivative with respect to $r$ . To see the behavior of the potential, we have plotted various $V_{eff}$ as a function of $r$. In Fig. \eqref{fig1}, we have especially wanted to reveal the effect of the rotating parameter $a$ on $V_{eff}$. As can be seen from Fig. \eqref{fig1}, the height of $V_{eff}$ barrier increase with the rising $a$ value. It may be useful to remind a well-known physical feature at this point: the tunneling probability depends on the energy of the incident particle relative to the height of the barrier. 

\section{Reflection and Transmission Coefficients}\label{sec5}

In this section, we shall study the wave scattering from the rotating polytropic black hole by employing the WKB method. As is well-known, Hawking radiation before passing the gravitational potential is different from one after passing the potential. This difference can be
measured by the so-called greybody factor \cite{SS4}, which is in particular related with the transmission coefficient through the black hole's effective potential \cite{SK4}. 

To obtain the reflection and transmission coefficients, let us first consider a propagating wave that reaches to the black hole ($r_{\ast}\rightarrow-\infty$) from the spatial infinity $r_{\ast}\rightarrow\infty$. Because of the effective potential, some of the waves are transmitted and thus the other parts are reflected. The scattering boundary conditions have the following form \cite{S5}
\begin{align}
\Psi &  =T\left(  \omega\right)  e^{-i\omega r_{\ast}}\text{ \ \ \ \ \ \ \ \ }%
r_{\ast}\rightarrow-\infty,\nonumber\\
\Psi &  =e^{-i\omega r_{\ast}}+R\left(  \omega\right)  e^{i\omega r_{\ast}%
}\text{ \ \ }r_{\ast}\rightarrow+\infty,\label{33}%
\end{align}

where $T\left(  \omega\right)  $ and $R\left(  \omega\right)  $ are transmission and reflection coefficients, respectively. To determine the square of the wave function's amplitude, we use the normalization condition:
\begin{equation}
\left\vert T\right\vert ^{2}+\left\vert R\right\vert ^{2}=1. \label{34}%
\end{equation}

The greybody factor is formulated as follows \cite{SK7}%
\begin{equation}
\gamma_{l}\left(  \omega\right)  =\left\vert T\left(  \omega\right)
\right\vert ^{2}.\label{35}%
\end{equation}

\begin{figure}[h]
\centering\includegraphics[width=18cm,height=11cm]{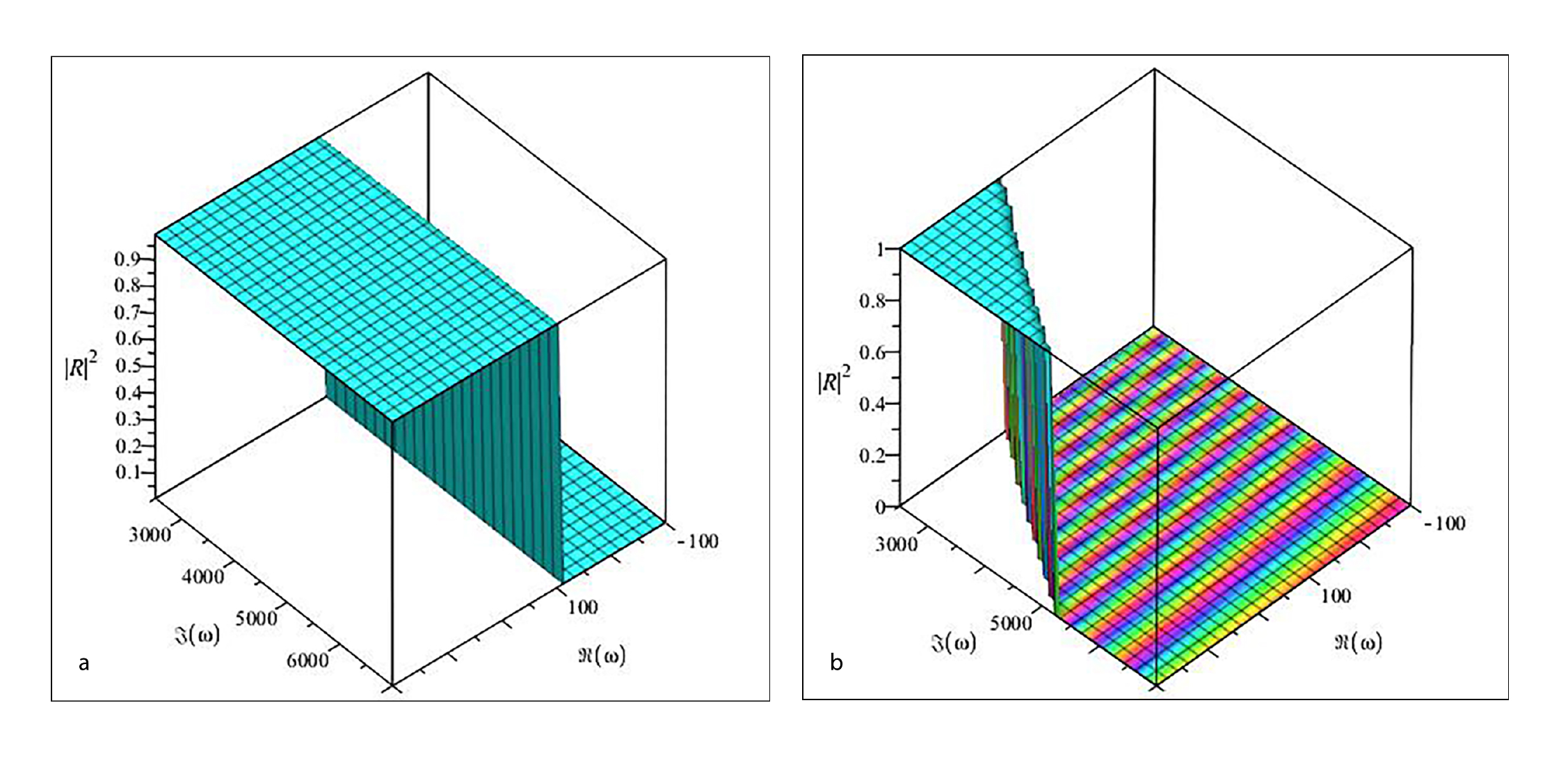}\caption{Plots
of reflection probability for $ a=0.011$ (\textit{\textbf{a}}) and $a=0.50$ (\textit{\textbf{b}}). The physical parameters are chosen as; $M=3,m=1,L=1$, and
$\lambda=6$.} \label{fig2}%
\end{figure}

One can follow the sixth order WKB formula \cite{S6}, which computes the reflection and transmission coefficients with the following expressions
\begin{equation}
R(\omega)=\frac{1}{\sqrt{1+\exp\left(  -2\pi i\mathfrak{K}\right)  }}, \label{36}%
\end{equation}

\begin{equation}
T(\omega)=\frac{1}{\sqrt{1+\exp\left(  2\pi i\mathfrak{K}\right)  }}, \label{37}%
\end{equation}
in which
\begin{equation}
\mathfrak{K}=i\frac{\omega_{n}^{2}-V\left(  r_{0}\right)  }{\sqrt{-2V^{\prime\prime
}\left(  r_{0}\right)  }}-\Lambda_{2}-\Lambda_{3}, \label{38}%
\end{equation}
where $r_{0}$ is the location of $r$ that makes the potential $V(r)$ maximum. From now on, we let $V(r_{0})=V_{0}$. Furthermore,
\begin{equation}
\Lambda_{2}=\frac{1}{\sqrt{-2V_{0}^{\prime\prime}}}\left[  \frac{1}{8}\left(
\frac{V_{0}^{\left(  4\right)  }}{V_{0}^{\prime\prime}}\right)  \left(
\frac{1}{4}+\alpha^{2}\right)  -\frac{1}{288}\left(  \frac{V_{0}^{\left(
3\right)  }}{V_{0}^{\prime\prime}}\right)  ^{2}\left(  7+60\alpha^{2}\right)
\right]  , \label{39}%
\end{equation}
and%
\begin{multline}
\Lambda_{3}=\frac{1}{\sqrt{-2V_{0}^{\prime\prime}}}\left[  \frac{5}%
{6912}\left(  \frac{V_{0}^{\left(  3\right)  }}{V_{0}^{\prime\prime}}\right)
^{4}\left(  77+188\alpha^{2}\right)  -\frac{1}{384}\left(  \frac{V_{0}%
^{\prime\prime\prime2}V_{0}^{\left(  4\right)  }}{V_{0}^{\prime\prime3}%
}\right)  \left(  51+100\alpha^{2}\right)  +\right. \\
\left.  \frac{1}{2304}\left(  \frac{V_{0}^{\left(  4\right)  }}{V_{0}%
^{\prime\prime}}\right)  ^{2}\left(  67+68\alpha^{2}\right)  -\frac{1}%
{288}\left(  \frac{V_{0}^{\prime\prime\prime}V_{0}^{\left(  5\right)  }}%
{V_{0}^{\prime\prime2}}\right)  \left(  19+28\alpha^{2}\right)  -\frac{1}%
{288}\left(  \frac{V_{0}^{\left(  6\right)  }}{V_{0}^{\prime\prime}}\right)
\left(  5+4\alpha^{2}\right)  \right]. \label{40}%
\end{multline}

In Eqs. \eqref{38}-\eqref{40}, both prime symbol and  superscript $\left(  4,5,6\right)$ denote the differentiation with respect to the tortoise coordinate $r_{\ast}$ and $\alpha=n+\frac{1}{2}$. 

\begin{figure}[h]
\centering\includegraphics[width=18cm,height=11cm]{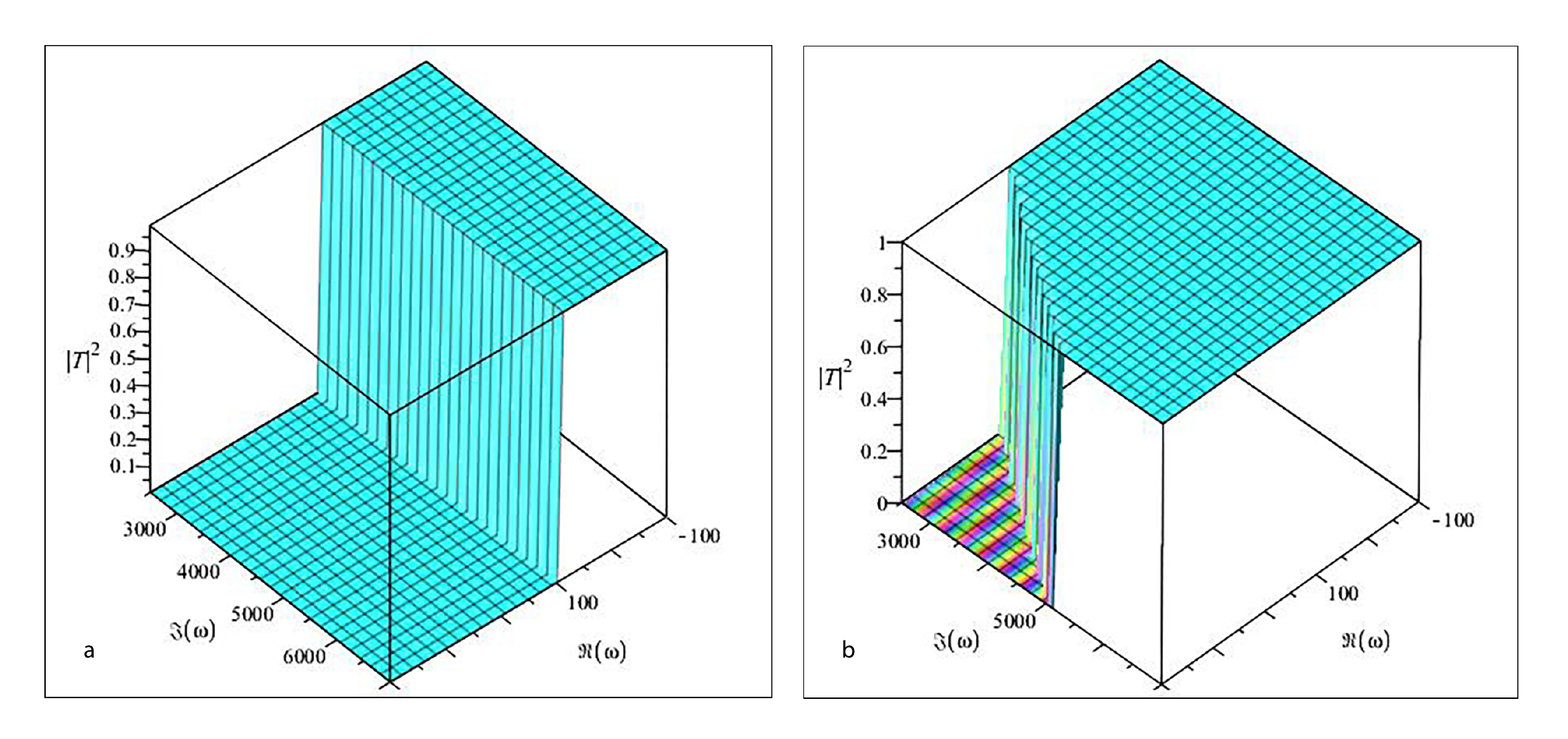}\caption{Plots
of transmission probability for $a=0.011$ (\textit{\textbf{a}}) and $a=0.5$ (\textit{\textbf{b}}). The physical parameters are chosen as; $M=3,m=1,L=1$, and
$\lambda=6$.} \label{fig3}%
\end{figure}

\begin{figure}[h]
\centering\includegraphics[width=15cm,height=11cm]{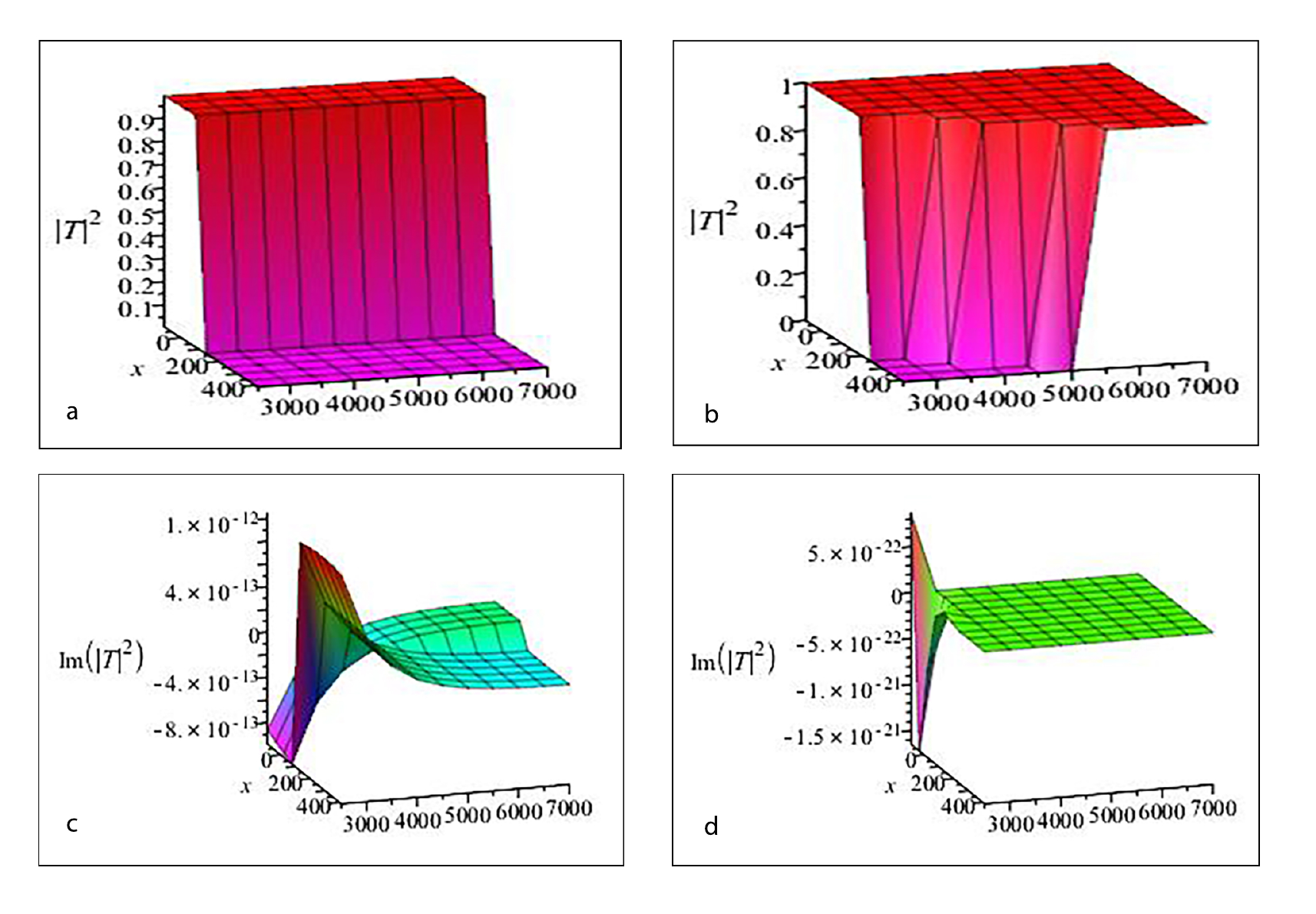}\caption{Comparison between the real (\textit{\textbf{a}} and \textit{\textbf{b}}) and imaginary (\textit{\textbf{c}} and \textit{\textbf{d}}) parts  of the transmission probability represented in Fig. \ref{fig3}}\label{fig4}%
\end{figure}

The transmission and reflection probabilities for the rotating parameter $ a=0.011$ and $a=0.5$ are depicted in Figs. \eqref{fig2} and \eqref{fig3}, respectively. In order to develop a broader perspective, Fig. \eqref{fig4} is depicted, which represents more detachments. Namely, under the change of $a$, the real and imaginary parts can be compared. It is seen that the imaginary part undergoes more changes than the real part with the variation of $a$ parameter.  

\section{Conclusions}\label{sec6}

In this article, we have considered scalar perturbations of a rotating polytropic black hole \cite{S1,K8} having a non-null cosmological parameter $L^{2}$. First, we have briefly the introduced the physical properties of the black hole and studied the quantum tunneling of the spin-$0$ particles from the considered black hole spacetime. We have derived the modified Hawking temperature by taking into account the GUP effects in our calculations. The obtained result \eqref{22} shows that GUP corrected temperature deviates from the standard Hawking temperature, which was corresponding to the black-body radiation in quantum gravity.

In the continuation of the study, we have analytically obtained the effective potential of the rotating polytropic black hole produced by the scalar perturbations. To this end, we have reduced radial part of the Klein-Gordon equation to the one-dimensional Schr\"{o}dinger like wave equation. The behaviors of the effective potential for the scalar waves traveling in this spacetime are depicted in Fig. \eqref{fig1}. We have deduced from Fig. \eqref{fig1} that the height of $V_{eff}$ barrier grows with the increasing of the rotating parameter $a$. In this study, the greybody factor $\gamma(\omega)$, which is equivalent to the transmission coefficient $|T(\omega)|^{2}$ [see Eq.\eqref{37} ] and also known as one of the basic keys of the information that can be obtained from the black hole, is studied. To this end, we have employed sixth order WKB approximation to compute the transmission and reflection coefficients, which are seen in Eqs.\eqref{36} and \eqref{37}.The greybody factor $\gamma(\omega)$ experiences growth  and hence the reflection coefficient $|R(\omega)|^{2}$ becomes larger with increase in $a$, whence, $|R(\omega)|$. On the other hand, we have found that the change in the rotation parameter $a$ does not radically change the the transmission and reflection coefficients.

In future, we plan to extend our current work to the fermionic, electromagnetic, and gravitational perturbations for the rotating polytropic black hole. By this way, we aim to reveal effect of the spin on the reflection and transmission coefficients with respect to the $L^2$ or to the cosmological constant parameter $\Lambda$.

\end{document}